\documentclass{jfm}

\usepackage{graphicx}
\usepackage{newtxtext}
\usepackage{xcolor}
\usepackage{newtxmath}
\usepackage{natbib}
\usepackage{hyperref}
\hypersetup{
    colorlinks = true,
    urlcolor   = blue,
    citecolor  = black,
}

\newcommand{\RomanNumeralCaps}[1]
\linenumbers

\usepackage{siunitx}
\usepackage{derivative}
\usepackage{booktabs} 
\usepackage[justification=justified,width=\textwidth]{caption}
\usepackage{upgreek}
\usepackage{tikz}
\usepackage{euscript}[mathcal]
\usepackage{bbding}

\DeclareMathOperator\erf{erf}
\newcommand{\diffratioL}{\sigma}

\definecolor{korr_blue}{rgb}{0.070588235,0.364705882,0.650980392}
\newcommand{\WS}[1]{#1}
\newcommand{\hollowstar}{\text{\FiveStarOpen}}

\title{Role of melting and solidification in the spreading of an impacting water drop}

\author{Wladimir Sarlin\aff{1}
 \corresp{\email{wladimir.sarlin@outlook.com}},
  Rodolphe Grivet\aff{1},
  Julien Xu\aff{1},
  Axel Huerre\aff{2},
  Thomas Séon\aff{3},
 \and Christophe Josserand\aff{1}
 \corresp{\email{christophe.josserand@ladhyx.polytechnique.fr}}}

\affiliation{\aff{1}Laboratoire d'Hydrodynamique, CNRS, École polytechnique, Institut Polytechnique de Paris, 91120 Palaiseau, France
\aff{2}Laboratoire Matière et Systèmes Complexes (MSC), UMR CNRS 7057, Université Paris Cité, 75013 Paris, France
\aff{3}Institut Franco-Argentin de Dynamique des Fluides pour l’Environnement (IFADyFE), IRL 2027, CNRS, UBA, CONICET, Buenos Aires, Argentine}

\begin{document}
\maketitle

\begin{abstract}
The present study reports experiments of water droplets impacting on ice or on a cold metal substrate, with the aim of understanding the \WS{effects of liquid solidification or substrate melting} on the impingement process. Both liquid and substrate temperatures are varied, as well as the height of fall of the droplet. The dimensionless maximum spreading diameter, $\beta_m$, is found to increase with both temperatures as well as with the impact velocity. Here $\beta_m$ is reduced when \WS{liquid} solidification, which enhances dissipation, is present, whereas fusion\WS{, \textit{i.e.}, substrate melting,} favours the spreading of the impacting droplet. These observations are rationalized by extending an existing model of effective viscosity, in which phase change alters the size and shape of the developing viscous boundary layer, thereby modifying the value of $\beta_m$. The use of this correction allows us to adapt a scaling recently developed in the context of isothermal drop impacts to propose a law giving the maximum diameter of an impacting water droplet in the presence of melting or solidification. \WS{Finally, additional experiments of dimethyl sulfoxide drop impacts onto a cold brass substrate have been performed and are also captured by the proposed modelling, generalizing our results to other fluids.}
\end{abstract}

\begin{keywords}
\end{keywords}


\section{Introduction}
\label{introduction}

In his poem \textit{De Rerum Natura}, Lucretius asks: ``\textit{don't you see, besides, how drops of water falling down against the stones at last bore through the stones?}''. This sentence, dating back to the first-century BC, is a testament to the long research interest for the problem of drop impacts on a substrate, which is always of topicality nowadays \citep{2016_josserand,2021_blanken,2022_cheng}. A better understanding of the droplet dynamics and maximum spreading diameter after impact is motivated by the wide range of industrial and natural applications such as, amongst others, spray deposition \citep{2002_pasandideh-fard}, aerosol generation \citep{2015_joung}, or raindrop erosion \citep{2015_zhao}. This led to the elaboration of models describing single drop impacts in the capillary and viscous limits \citep{2010_eggers}, or in the transition between these two asymptotic regimes \citep{2014_laan,2016_lee}, \WS{in an isothermal context (\textit{i.e.}, when no thermal effects are involved)}.

\WS{The particular configuration of drop impact involving phase change also received a significant attention due to its relevance for three-dimensional or inkjet printing \citep{2016_wang,2022_lohse}, spray coating or cooling processes \citep{2015_shukla,2018_breitenbach}, or aircraft icing problematics \citep{2018_baumert}, for instance. Several studies focused on droplets impinging on heated walls \citep{2010_moita,2017_liang,2013_quere}, and identified different regimes for the spreading dynamics. In particular, \citet{1991_chandra} explored experimentally the case of $n$-heptane droplets impinging at a low impact velocity on a stainless steel substrate whose surface temperature could be varied from 24 \textcelsius\ to 250 \textcelsius, encompassing both the liquid boiling point and the Leidenfrost point. This allowed these authors to describe the spreading process below and above the Leidenfrost point. In the second situation, the impacting droplet levitates above the substrate, due to the formation of a vapour layer under the expanding liquid film. From there, \citet{2012_tran} investigated water drop impacts on hot surfaces, and provided a comprehensive phase diagram highlighting the existence of three regimes: contact boiling, gentle film boiling, and spraying film boiling. \citet{2015_staat} studied the impact of ethanol droplets on a hot surface, varying both the Weber number and the substrate temperature, to determine the transition towards splashing and the dynamic Leidenfrost point (onset of the Leidenfrost effect). They evidenced a strong dependency of the splashing threshold with the substrate temperature. The transition regime between contact boiling and film boiling has been investigated by \citet{2016_shirota} using total internal reflection imaging. Another experimental contribution from \citet{2013_antonini} revealed that the Leidenfrost effect, superhydrophobicity, and sublimation of the substrate have a similar influence on the dynamics of an impinging drop, with droplet rebound being observed in each situation. \citet{2020_liu} recently studied drop impacts on heated nanostructures, and highlighted that hot nanotextures can enhance jetting and splashing during the impact process.}

\WS{Although the case of drop impact on a hot wall has been the subject of important scientific literature, fewer experimental studies have been dedicated to situations featuring liquid solidification or substrate melting. In a seminal contribution, \citet{1976_madejski} derived a theoretical analysis of the spreading dynamics of a liquid droplet impinging onto a solid substrate cold enough to trigger solidification, based on energy conservation, alongside experiments of metal drop impacts performed on different substrates. Numerous studies have since been dedicated to the case of a metal drop impinging on a cold substrate, providing estimates for the maximum spreading diameter \citep{1998_pasandideh-fard,2020_gielen}, or evidencing the intriguing self-peeling phenomenon occurring for a cold enough surface temperature \citep{2018_de_ruiter}. When studying the outcome of water droplet impacts on a cold substrate, \citet{2016_ghabache} observed different crack patterns developing in the resulting frozen puddle, depending on the surface temperature, and proposed a model to estimate the thresholds towards the fragmentation and the hierarchical regimes. \citet{2018_schremb} studied the impact of supercooled water droplets on a smooth ice target, and developed an analytical framework to describe the lamella thinning as well as the final ice thickness. Recently, \cite{2020b_thievenaz} studied the influence of solidification (or freezing) on the maximum spreading of a water droplet impacting on a cold surface, at rather large impact velocities. These authors proposed a model of effective viscosity which allowed them to successfully describe their experiments. If these studies focused on the sole case of solidification during a drop impact, recent experimental works investigated the thermodynamic configuration in which the droplet is able to melt the solid surface it impacts \citep{2017_jin,2019_ju,2022_lolla}, but a model describing the effect of substrate melting on the impact outcome is currently missing.}

\WS{Therefore, a unified description of the influence of phase change on the maximum spreading diameter resulting from a drop impact remains elusive. In particular, although these situations are conceptually close, no modelling describes the effects of substrate melting and liquid solidification on the impact outcome in a common framework.} These aspects motivated the present experimental study, which aims at investigating the case of a temperature-controlled water droplet falling onto a cold substrate, made either of ice or of cold brass, in order to understand the effects of substrate melting and liquid solidification on the maximum spreading diameter. 

\section{Experimental set-up and methods}
\label{setup}

\begin{figure}
    \centerline{
    \includegraphics[width=\linewidth]{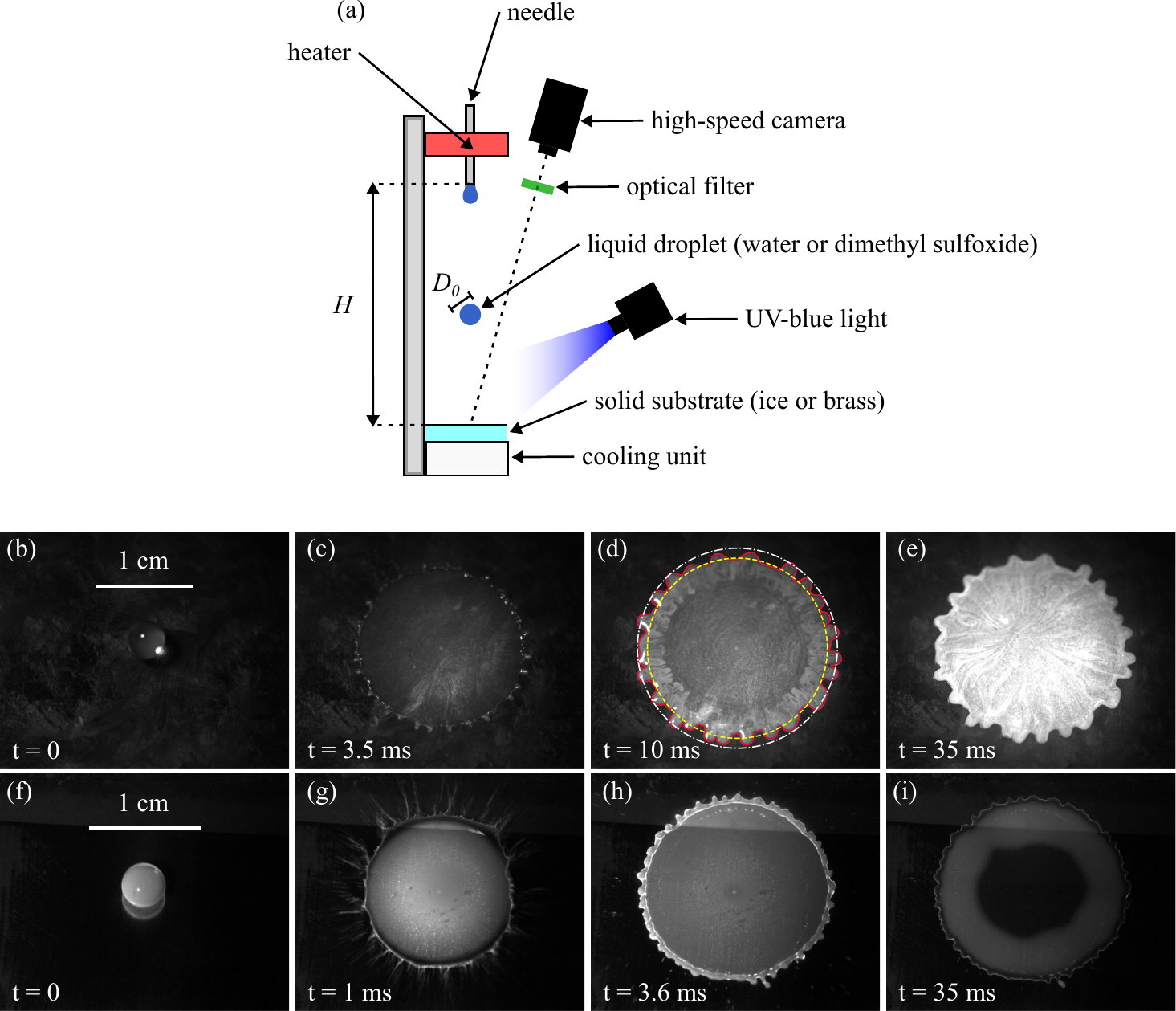}
    }
    \caption{(a) Schematic representation of the experimental set-up. The initial diameter of the liquid droplet is $D_0$, and the distance between its original position and the substrate is $H$. (b)-(e) Photographs of a transparent water droplet spreading on melting tinted ice, with $T_s=-5$ \textcelsius, $T_d=25$ \textcelsius, and $H=0.4$ m, at times (b) $t=0$, (c) $t=3.5$ ms, (d) $t=10$ ms, and (e) $t=35$ ms after impact. (f)-(i) Pictures of a tinted water droplet spreading on cold brass, with $T_s=-32.2$ \textcelsius, $T_d=19.3$ \textcelsius, and $H=1.5$ m, at times (f) $t=0$, (g) $t=1$ ms, (h) $t=3.6$ ms and (i) $t=35$ ms after impact. In (b)-(e) and (f)-(i), melting and solidification are evidenced by the increase or decrease of the brightness with time, respectively, which reveals that more and more ice melts (respectively, an increasing amount of water solidifies). In (d), the red line is the extracted contour of the liquid film when it reaches its maximum radial extent. Circles of diameter $D_{\rm{min}}$ (yellow dashed line) and $D_{\rm{max}}$ (white dash-dotted line) \WS{corresponding to the averaged minimum and maximum droplet diameter, respectively,} are also represented. Horizontal bars in (b) and (f) give the scales for each corresponding image sequence.}
    \label{setup_pictures}
\end{figure}

The experimental set-up designed to this end is schematized in Figure \ref{setup_pictures}(a). It consists of a $2.2\ \rm{m}$ high vertical beam, that holds an aluminium block that can be set at an adjustable height. A vertical needle of \WS{outer diameter $1.83\ \rm{mm}$} passes through the block, which was hollowed out to host a heating cartridge placed in contact with the needle and connected to a generator. At the bottom of the beam, a cooling unit can be used as is, or to generate an ice layer: it is made of a thin brass plate, cooled by a Peltier heat sink that is in turn connected to a cold bath operating with a mixture of water and ethylene glycol. Drop impacts can then be made directly onto cold brass, to study the role of liquid solidification during the impingement process, or onto an ice layer, which is produced by depositing a certain amount of water on the plate before freezing it rapidly with the help of the Peltier modulus. When conducting experiments with the brass substrate, a perspex plate is placed on top of it to limit the formation of frost, without altering much its surface temperature due to the low thermal effusivity of plastic. This protection is removed just before performing a drop impact. A thermocouple, placed on top of the substrate, allows us to adjust the heat flux imposed by the Peltier so that the brass or ice layer is set at an initial surface temperature $T_s$. At the beginning of an experiment, the needle is positioned so that its tip is located at a distance $H$ from the substrate (with a $4$ mm accuracy). Then, the liquid is gently pushed into the needle using a syringe pump, and a pendant drop is formed. \WS{The droplet temperature is set and maintained at a controlled value $T_d$ using a regulation loop, based on a tension generator connected to the heating cartridge and to two thermocouples, that are respectively placed inside the needle and in contact with the cartridge. As long as the temperature in the needle is lower than $T_d$, the cartridge is powered and heats the liquid up, but it is switched off when the measured temperature becomes greater than or equal to $T_d$. This regulation system allows us to quickly obtain a very stable initial temperature for the pendant drop, which is measured with a $\pm 2$ \textcelsius\ accuracy. The syringe pump is then activated anew to slowly inject liquid inside the drop so that it eventually reaches its critical volume and detaches from the needle under the action of gravity. As a result, a droplet of initial diameter $D_0$ starts its fall over the vertical distance $H$ before impacting and spreading over the ice or brass substrate.} This process is recorded from above by a high-speed camera, which operates at 5000 fps.

\WS{In most experiments reported in the present study, the liquid used is pure water, whose initial temperature has been varied between $18$ \textcelsius\ and $80$ \textcelsius, while the substrate temperature $T_s$ ranged between $-33$ \textcelsius\ and $-2$ \textcelsius\ for both the ice or brass surface. The height $H$ has been explored in the range $\left[ 0.04 - 2.2 \right]$ m so that the resultant impact velocity $U$ of the water droplet, which is evaluated by a home-made code accounting for the air resistance, is varied between $0.9\ \rm{m.s^{-1}}$ and $5.9\ \rm{m.s^{-1}}$.}

\WS{A complementary set of experiments of dimethyl sulfoxide (DMSO) drop impacts on cold brass have also been performed, as this fluid has significantly different thermal properties than water, and also a reduced surface tension. For these particular tests, the droplet temperature was kept constant (at $T_d=25$ \textcelsius), whereas three substrate temperatures ($T_s=-30$ \textcelsius, $T_s=10$ \textcelsius, and $T_s=30$ \textcelsius) and four initial heights ($H=10$ cm, $H=50$ cm, $H=128$ cm, and $H=170$ cm) have been investigated. The explored surface temperatures have been chosen so as to have one temperature above the DMSO freezing point (which is $T_f \simeq 18.6$ \textcelsius), and two below. As a result, these experiments solely involve no phase change or liquid solidification.}

\WS{For each initial condition, experiments are repeated three times to ensure reproducibility. Using standard correlations for the convective heat transfer of a sphere moving in a fluid \citep{1960_yuge}, the maximum temperature drop during the fall of a water droplet can be estimated to be less than 2 \textcelsius\ at worst, \textit{i.e.}, when $T_d=80$ \textcelsius\ and $H=2.2$ m, as described in Appendix \ref{appendix_B}. In the case of experiments with DMSO, for which $T_d$ is very close to the ambient air temperature, the estimated change is below 1 \textcelsius. As a result, we choose to neglect this effect so that the droplet temperature at the time of impact is considered to be equal to $T_d$. Finally, the initial droplet diameter $D_0$ has been measured for all experiments from the last image showing the drop before collision with the substrate: it is found to be $D_0 \simeq 4.0 \pm 0.15\ \rm{mm}$ for water and $D_0 \simeq 3.2 \pm 0.15\ \rm{mm}$ for DMSO.}

\section{Qualitative and quantitative results}
\label{results}

In the present experiments, when a liquid droplet impacts the substrate, it starts spreading radially on top of it. This dynamics happens on the characteristic kinetic time scale $D_0/U$, of the order of milliseconds in the present experiments, until the drop reaches its maximum spreading diameter. After this moment, there is no significant retraction of the contact line on the substrate. 

An important aspect is to determine whether the ice effectively melts on the same time scale when a hot water droplet impacts on its surface, or if the liquid (water or DMSO) solidifies when impinging a cold brass substrate. To this end, two kinds of preliminary experiments are performed with water droplets, where either the solid or liquid phase is dyed with fluorescein. The spreading is illuminated from above using a UV-blue light (with a wavelength of $470\ \rm{nm}$), and the camera lens is covered by a green optical filter (with cutting wavelength of  $495\ \rm{nm}$), so that only fluorescent regions appear bright on the obtained images. In the first situation, an ice layer is dyed during its formation, whereas the impinging drop remains translucent. As fluorescence does not happen when fluorescein molecules are diluted in solid water due to a self-quenching phenomenon \citep{2021_huerre}, its detection is a signature of melted water originating from the ice layer. \WS{The pictures in Figure \ref{setup_pictures}(b)-(e), corresponding to an experiment for which $T_s=-5$ \textcelsius, $T_d=25$ \textcelsius, and $H=0.4$ m, show (b) the drop prior to the collision with the substrate, (c) the liquid film during the spreading process, (d) the moment the maximum diameter is reached, and (e) the final footprint left by the impact.} If the initial droplet is almost not visible due to its transparency, the liquid phase becomes increasingly luminous as time goes by: this demonstrates the melting of the ice during the whole impact process. Conversely, in the second case corresponding to a water drop impact onto cold brass, only the impinging liquid has been dyed. \WS{Figure \ref{setup_pictures}(f)-(i) presents a typical image sequence of this configuration for an experiment where $T_s=-32.2$ \textcelsius, $T_d=19.3$ \textcelsius, and $H=1.5$ m, with (f) the tinted droplet just before the first contact with the substrate and (i) the resulting imprint long after the spreading phase.} The fact that the intensity in the liquid decreases with time, with some part of the expanding droplet becoming increasingly dark, reveals that solidification of the liquid layer is at play. It should be underlined that these phenomena are also visible during spreading: this suggests that the two typical times of radial expansion and of phase change are of the same order, so that there is no scale separation between the two processes. \WS{Another comment arises from the images of Figure \ref{setup_pictures}: the centre of the spreading water droplet pictured in Figure 1(h), photographed at the moment the maximum diameter has been reached for the corresponding experiment, remains fluorescent (hence, some liquid remains) and not completely dark (hence, solid) as in Figure 1(i). This shows that the arrest criterion does not correspond to the moment the liquid-solid interface is reached by the droplet free surface.}

In order to study quantitatively the impact outcome, we performed drop impact experiments on ice or cold brass, in which only the liquid (water or DMSO) has been tinted with fluorescein and the parameters $T_d$, $T_s$, and $H$ have been varied systematically in the ranges indicated in section \ref{setup}. \WS{In what follows, we specifically focus on the maximum diameter reached by the spreading liquid film that, under the presence of solidification or substrate melting, has received little attention so far in the literature. It should be mentioned, though, that the transient dynamics has already been addressed by several previous studies \citep{2017_jin,2019_ju,2020b_thievenaz}, showing that the spreading dynamics itself was only smoothly affected by the phase change.}

The final contour \WS{of the contact line}, obtained when the liquid film reaches its maximum radial extent, can then be extracted as illustrated by the red solid line in Figure \ref{setup_pictures}(d). Image processing allows us to locate the positions of the local maxima (\textit{i.e.}, the tip of the digitations) and minima (located between two fingers), relative to the centre of mass of the contour ($O$). From the corresponding radial distances to $O$, the minimum and maximum \WS{droplet} diameters, noted respectively as $D_{\rm{min}}$ and $D_{\rm{max}}$, are defined as the averaged positions of the local minima and maxima, respectively. From all our experiments, we observe a linear relationship between $D_{\rm{max}}$ with $D_{\rm{min}}$ regardless of the droplet or substrate temperatures, which reads $D_{\rm{max}}=1.07D_{\rm{min}}$ \WS{for water and $D_{\rm{max}}=1.04D_{\rm{min}}$ for DMSO.} As a result, the more the spreading, the more the fingers' elongation. The proportionality between the two diameters is an intriguing result, which suggests, for instance, that for water, the typical size of the digitations $D_{\rm{max}}-D_{\rm{min}}$ is about $7\%$ of the spreading diameter $D_{\rm{min}}$, and that $D_{\rm{max}}$ can be described in a similar way as $D_{\rm{min}}$. As a result, the maximum spreading ratio is defined as $\beta_m \equiv D_{\rm{min}}/D_0$, with $D_0$ the initial diameter of the droplet.

\begin{figure}
    \centerline{
    \includegraphics[width=\linewidth]{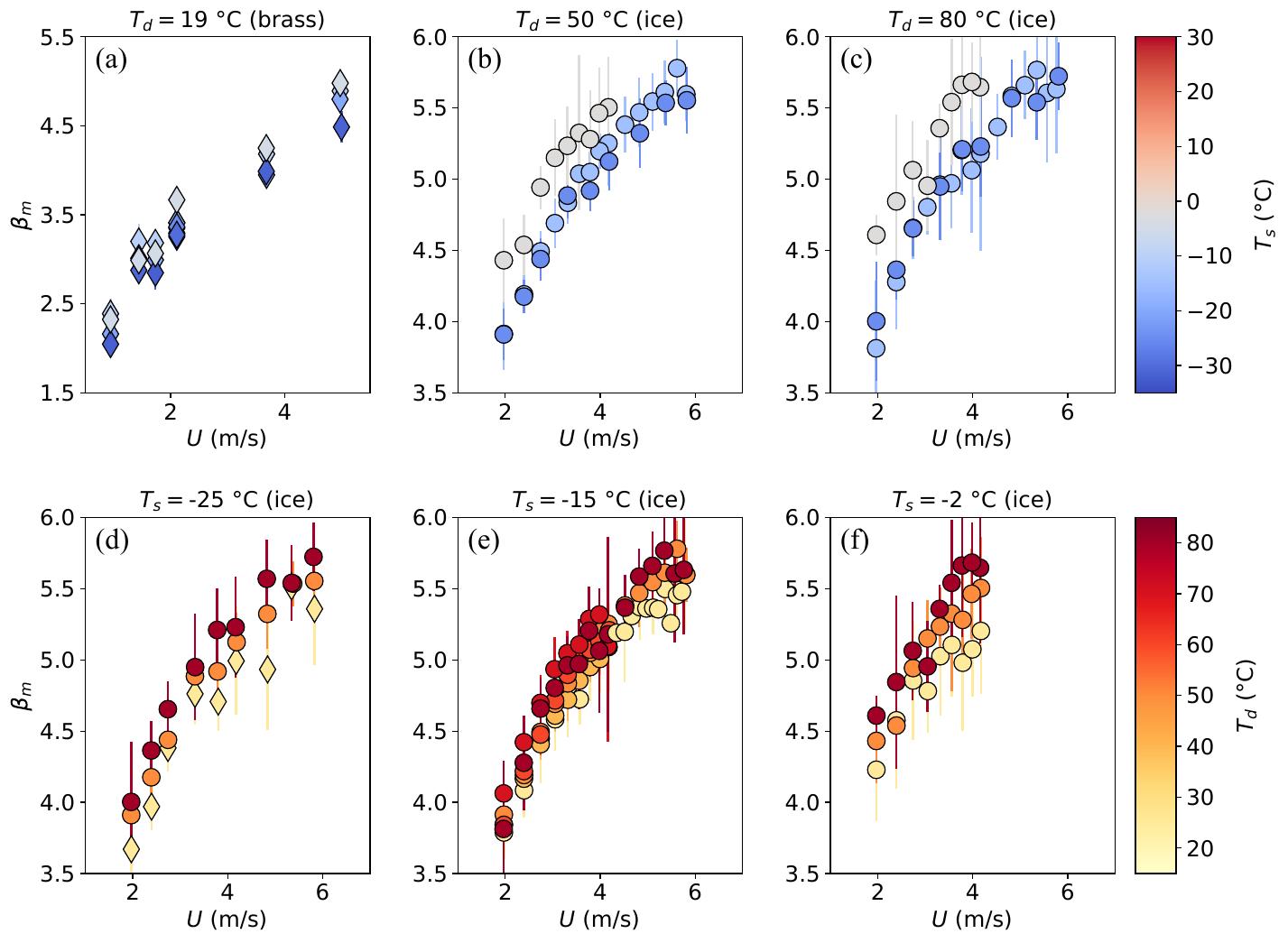}
    }
    \caption{Evolution of the maximum spreading ratio, $\beta_m \equiv D_{\rm{min}}/D_0$, as a function of the impact velocity $U$\WS{, for water drop impacts on (a) brass and (b)-(f) on ice}. In (a)-(c), the droplet temperature is fixed at (a) $T_d=19$ \textcelsius, (b) $T_d=50$ \textcelsius, and (c) $T_d=80$ \textcelsius, while the colourbar denotes the substrate temperature $T_s$. Contrariwise, in (d)-(f) the substrate temperature is fixed at (d) $T_s=-25$ \textcelsius, (e) $T_s=-15$ \textcelsius, and (f) $T_s=-2$ \textcelsius, while the markers' colours represent this time the droplet temperature $T_d$. The symbols correspond to experiments with water where ($\lozenge$) solidification or ($\medcirc$) fusion occurs, respectively. The nature of the substrate is indicated above each plot.}
    \label{betam_vs_U}
\end{figure}

In Figure \ref{betam_vs_U}, $\beta_m$ is presented as a function of the impact velocity $U$ \WS{for experiments involving water with} a fixed (a)-(c) droplet temperature $T_d$ or (d)-(f) substrate temperature $T_s$, which are indicated above each plot. \WS{The markers correspond to experiments involving water, and for which ($\lozenge$) solidification or ($\medcirc$) fusion (substrate melting) occurs, respectively. As it was not always straightforward to distinguish which kind of phase change was at play for a given experiment, the marker to assign is determined by the sign of the moving ice-liquid front position, which is calculated from the results presented in section \ref{discussion}.} In all cases, the spreading ratio clearly increases with $U$. Furthermore, for a fixed value of $T_d$, increasing $T_s$ results in larger $\beta_m$: the higher the substrate temperature, the larger the spreading ratio. This is illustrated for ice in (b) and (c) where $T_d=50$ \textcelsius\ and $T_d=80$ \textcelsius, respectively: in these situations, experiments conducted at $T_s=-2$ \textcelsius\ (light grey) are significantly above those performed at $T_s=-25$ \textcelsius\ (blue). To a lesser extent, at a given value of $T_s$, $\beta_m$ is larger when $T_d$ is increased. This is especially visible in (f) for $T_s=-2$ \textcelsius, where a gentle order exists with the value of the droplet temperature $T_d$. These observations agree with the results gathered by \citet{2017_jin} and \citet{2019_ju} for drop impacts on ice, and by \cite{2020b_thievenaz} for droplet impingement on a cold metal substrate. \WS{Experiments with DMSO present a similar evolution as the water drop impact tests presented in figure \ref{betam_vs_U}(a). They are not included here, as they feature a slightly larger initial droplet temperature ($T_d=25$ \textcelsius).}

\section{Discussion and modelling}
\label{discussion}

\WS{Building upon previous studies dedicated to drop impacts \citep{1976_madejski,2010_eggers,2012_lagubeau,2014_laan,2016_lee,2016_josserand}, two dimensionless numbers can be defined to describe the outcome of the impact process. On one hand, the Weber number $\rm{We} \equiv \rho U^2 D_0 / \gamma$ compares inertia to capillarity, with $\rho$ and $\gamma$ the density and surface tension of the drop, respectively. The values for $\rho$ and $\gamma$ involved in the expression of $\rm{We}$ are taken here at $T_d$ using standard correlations (see Appendix \ref{appendix_A}). This choice was made as both the initial kinetic energy of the droplet and the surface energy once the maximum diameter has been reached are expected to involve liquid volume (respectively, surface) set at that temperature. Indeed, one can show that the thermal boundary layer in the liquid is always much smaller than the liquid film thickness when the maximal diameter is reached. On the other hand, the Reynolds number, defined here as $\rm{Re} \equiv U D_0/\nu_f$, with $\nu_f$ the kinematic viscosity, compares inertia to viscous effects. This time, $\nu_f$ is evaluated at the melting point $T_f$ since, in the spreading dynamics, this is the viscosity close to the substrate (thus, near $T_f$) that is relevant. Although this might seem surprising, at first glance, as the fluid properties for $\rm{We}$ have been evaluated at $T_d$, taking the kinematic viscosity at the initial temperature of the droplet, $T_d$, resulted in a significantly enhanced scattering of our results, which clearly suggests that the melting point is more relevant to describe the typical temperature of the dissipative layer for our experiments.} 

\begin{figure}
    \centerline{
    \includegraphics[width=\linewidth]{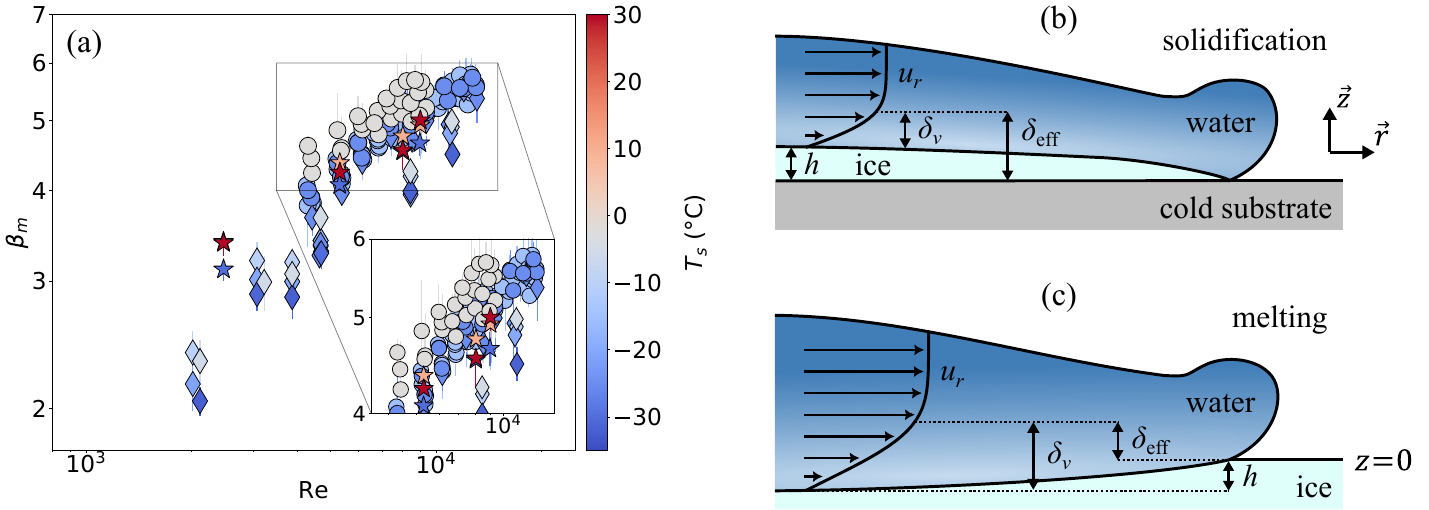}
    }
    \caption{(a) Maximum spreading ratio, $\beta_m$, as a function of the Reynolds number, $\rm{Re}$. The colourbar indicates the substrate temperature, while the symbols denote experiments involving ($\lozenge$) water droplets experiencing freezing, ($\medcirc$) water droplets causing ice melting, and (\scalebox{0.85}{$\hollowstar$}) DMSO droplets. (b)-(c) Schematic views of the ice (b) growth or (c) melting during film spreading. Here, $\delta_\nu$ corresponds to the size of the viscous boundary layer, $h$ is the position of the substrate, and $\delta_{\rm{eff}}$ the size of the effective boundary layer. $u_r$ is the radial velocity field.}
    \label{betam_vs_Re}
\end{figure}

\WS{Except for a few tests that have a small falling distance $H$, most of the impact velocities in the present experiments are greater than or equal to $2\ \rm{m.s^{-1}}$. For these data $230 < \rm{We} < 2000$ and $4500 < \rm{Re} < 13100$: as $\rm{Re} \gg 100$, the impact outcome is thus expected to be closer to the inertial-viscous regime provided by \citet{1976_madejski} than to the inertial-capillary regime, and hence, our data to be relatively well parameterised by the Reynolds number.} The evolution of $\beta_m$ with $\rm{Re}$ is illustrated in Figure \ref{betam_vs_Re}(a). Overall, the spreading ratio is found to increase with the Reynolds number, but it can be noted that the data are scattered in this representation. Indeed, an order appears with $T_s$, which is visible, for instance, for water drop impacts in which ice melts ($\medcirc$), as illustrated by the inset of Figure \ref{betam_vs_Re}(a). In addition, for a given Reynolds number, experiments featuring water solidification ($\lozenge$) have a systematically smaller value for $\beta_m$ than data involving ice melting ($\medcirc$). \WS{Last but not least, DMSO drop impacts (\scalebox{0.85}{$\hollowstar$}), featuring either no phase change or liquid solidification, lie systematically above experiments with water solidification ($\lozenge$).}

The poor collapse of the data in this representation is, in fact, expected as the spreading dynamics is affected by the presence of phase change. \WS{In the case of isothermal drop impacts belonging to the inertial-viscous regime, it has been shown that the arrest criterion corresponds to the moment the viscous boundary layer reaches the free surface of the expanding liquid film \citep{2010_eggers}. Building upon this, \citet{2020b_thievenaz} evidenced the fact that, for experiments involving solidification, spreading appears to stop when the sum of the growing ice thickness and the developing viscous boundary layer reaches the free-surface elevation of the expanding droplet. Further possible evidence of such an arrest criterion may be found in the study performed by \cite{1998_pasandideh-fard}, who studied tin drop impacts onto a cold stainless steel substrate. Indeed, in the numerical simulations conducted by these authors, it can be observed that spreading stops when the solidified layer at the centre of the splat approaches the free surface, whereas other regions of the spreading film remain in the liquid state.}

\WS{These results from previous studies shed light on the relevance to predict the moment the viscous boundary layer reaches the free surface of the liquid film when solidification or substrate melting occurs.} In such a situation, the ice grows or melts at the base of the expanding liquid film, as illustrated in Figure \ref{betam_vs_Re}(b)-(c), thus changing the position of the solid surface on which the viscous boundary layer of thickness $\delta_{\nu}$ develops, thereby modifying the value of $\beta_m$. The position of the moving interface can, at first order, be modelled by solving the classical Stefan problem, if one assumes that the influence of advection is negligible. The importance of this latter effect is known to depend on the Prandtl number $\rm{Pr}$, which compares the typical sizes of the thermal and viscous boundary layers \citep{2010_roisman}. As in the case of water close to its melting point this number is quite large ($\rm{Pr} \sim 14$ at $0.01$\textcelsius), advection can safely be neglected in the absence of phase change \citep{2010_moita}. We assume that the presence of solidification or melting does not change much this result much. \WS{Then, the position of the moving interface can be estimated by solving the heat diffusion equations in all phases (liquid, ice, and possible substrate) alongside the so-called ``Stefan condition'', which states that the velocity of the phase change front is directly related to the thermal flux difference at the boundary. The calculation leads to the derivation of a self-similar solution for the temperature field and the liquid-ice front position (see Appendix \ref{appendix_C} for more details)}. To put it in a nutshell, under these assumptions the front of the ice layer $h$ follows a diffusive law of the form $h(t)= s \sqrt{\alpha_{\rm{eff}}t}$, where $s=-1$ in the case of melting (respectively, $s=1$ for solidification), and $\alpha_{\rm{eff}}$ is an effective thermal diffusivity ($\alpha_{\rm{eff}} \geqslant 0$). By introducing $\chi=s\sqrt{\alpha_{\rm{eff}}/\alpha_i}$, with $\alpha_i$ the ice thermal diffusivity, $\alpha_{\rm{eff}}$ and $s$ are found numerically by solving the transcendental equation on $\chi$,

\begin{equation}
    \label{transcendental_equation}
    \frac{\chi \sqrt{\pi}}{2 \rm{St}} = \frac{e^{-\chi^2/4}}{r_i/r_s+\erf \left( \chi/2 \right)} + \frac{r_d}{r_i} \frac{e^{-\chi^2/(4 \omega_d)}}{1-\erf \left( \chi/(2\sqrt{\omega_d}) \right)} \frac{T_f-T_d}{T_f-T_s},
\end{equation}

\noindent with $r_d$, $r_i$ and $r_s$ the thermal effusivities of the liquid, the ice, and the possible substrate, respectively; $\omega_d=\alpha_d/\alpha_i$, with $\alpha_d$ the liquid thermal diffusivity; and $\mathrm{St}=c_{p,i} (T_f-T_s) / \mathcal{L}_f$ the Stefan number, with $c_{p,i}$ the ice thermal capacity and $\mathcal{L}_f$ the latent heat of fusion \WS{(for the definitions of these quantities, see also Appendix \ref{appendix_A})}. \WS{For drop impacts on ice, it should be noted that $r_s=r_i$. For each experiment, the value for $s$ obtained when solving equation (\ref{transcendental_equation}) indicates whether freezing ($s=1$) or melting ($s=-1$) occurred, so that the symbols used in Figures \ref{betam_vs_U}, \ref{betam_vs_Re}(a), and \ref{betam_vs_Reeff_Pade_eff} are chosen accordingly.}

From there, following the approaches developed by \citet{2010_eggers}, \citet{2010_roisman} and later by \citet{2020b_thievenaz}, it is possible to estimate the size of the viscous boundary layer relative to the initial position of the substrate, which is expected to eventually dictate the arrest. This is done by considering the $r$ component of the axisymmetric Navier-Stokes equations, in the Prandtl boundary layer framework for an incompressible flow:

\begin{equation}
    \label{axi_ns}
    \partial_t u_r + u_r \partial_r u_r + u_z \partial_z u_r = \nu_f \partial_z^2 u_r.
\end{equation}

\noindent Here $u_r$ and $u_z$ are the radial and vertical components of the velocity field, respectively, and $\partial_a$ stands for partial differentiation with respect to variable $a$. In the inviscid case and in the absence of phase change, using the streamfunction $\psi$ defined from $u_r \equiv - \partial_z \psi/r$ and $u_z \equiv \partial_r \psi /r$, the solution describing the impact can be taken as $\psi=-r^2z/t$, corresponding to a time decreasing arrest point flow with $u_r=r/t$ and $u_z=-2z/t$. Then, in the situation of a viscous flow subjected to solidification or melting, since both the viscous boundary layer, growing from $z=h(t)$, and the solid-liquid front position $h(t)$ follow a diffusive-in-time evolution, we can consider the following ansatz for the streamfunction:

\begin{equation}
\psi \equiv \sqrt{\nu_f} \frac{r^2}{\sqrt{t}} f ( \zeta ),
\label{ansatz}
\end{equation}

\noindent where $\zeta=[z-h(t)]/\sqrt{\nu_f t}$ is the self-similar variable and $f$ an unknown function of $\zeta$. As $u_r=-(r/t) f^\prime (\zeta)$, $f^\prime$ provides an insightful description of the shape of the boundary layer. Following \citet{2020b_thievenaz}, we inject the expression of $\psi$ into equation \eqref{axi_ns}, which leads to

\begin{equation}
    \label{self_similar_equation}
    f''' = -f' - \frac{1}{2} \left( \zeta + s \sqrt{\frac{\alpha_{\rm{eff}}}{\nu_f}} \right) f'' - f'^2 + 2ff''.
\end{equation}

\begin{figure}
    \centerline{
    \includegraphics[width=\textwidth]{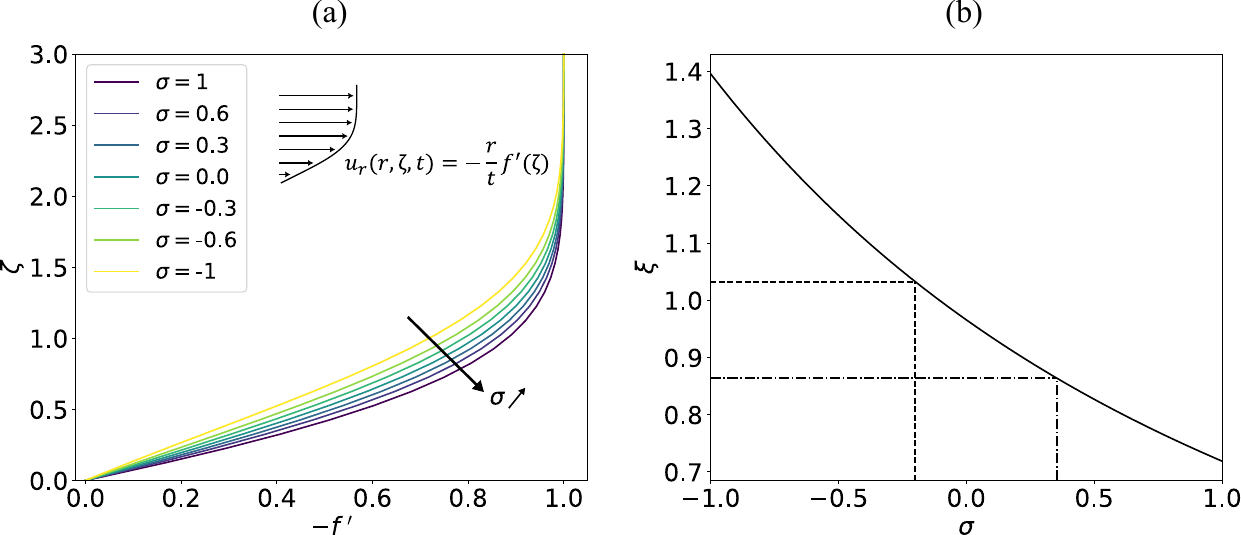}
    }
    \caption{(a) Evolution of the self-similar variable $\zeta=[ z - h(t) ]/\sqrt{\nu_f t}$ as a function of $-f^\prime$ (``velocity profile'' representation). Each curve corresponds to a given value of $\diffratioL \equiv s \sqrt{\alpha_{\rm{eff}}/\nu_f}$. (b) Evolution of the form factor $\xi=-1/f''(0)$ of the viscous boundary layer as a function of $\diffratioL$. The dashed and dash-dotted lines highlight the upper and lower limits for $\xi$ covered in the present experiments, respectively.
    }
    \label{blc}
\end{figure}

\noindent The boundary conditions are a zero velocity at the solid-liquid interface $\zeta=0$, and the recovery of the inviscid profile at infinity: this translates into $f(0)=0$, $f'(0)=0$, and $f'(+\infty)=-1$, respectively. The resolution of equation \eqref{self_similar_equation} for these boundary conditions is achieved numerically, by using a shooting method algorithm. The evolution of $\zeta$ as a function of $-f^\prime$ (\textit{i.e.}, in a ``velocity profile''-like representation) is illustrated in Figure \ref{blc}(a), for several representative values of $\diffratioL \equiv s \sqrt{\alpha_{\rm{eff}}/\nu_f}$. For each case, $-f^\prime$ increases from $0$ at the contact with the substrate to $1$ for $\zeta \simeq 2$, where the inviscid flow solution is thereby recovered. Furthermore, the curves for different values of $\diffratioL$ depart from each other, with those for large values of this parameter reaching the asymptotic behaviour earlier, meaning that the viscous boundary layer in this case is reduced in size when compared with lower $\diffratioL$. Therefore, this shows that the typical form factor of the viscous boundary layer, which can roughly be estimated as $\xi \simeq -1/f^{\prime \prime}(0)$, is a function of $\diffratioL$. In other terms, the present modelling predicts a coupling between the flow and the phase change dynamics. This fact can be verified in Figure \ref{blc}(b), where $\xi$ decreases with $\diffratioL$ in a weakly nonlinear manner. In the present experiments, $\xi$ ranges from 0.86 (for $\diffratioL \simeq 0.35$) to 1.04 (for $\diffratioL \simeq -0.2$). These values are highlighted in Figure \ref{blc}(b) by the dash-dotted and dashed lines, respectively.

It should be mentioned that a zero velocity condition has been imposed at the phase change front in the above analysis, although a volume-change flow actually exists at the liquid-ice interface during solidification or melting. However, a rough estimate of the induced velocity $v_{\rm{pc}}$ gives $v_{\rm{pc}} \simeq (\Delta \rho/\rho_d) \frac{\mathrm{d}h}{\mathrm{d}t}$, with $\Delta \rho = \rho_d - \rho_i$ ($\rho_d$ and $\rho_i$ being the liquid and ice densities taken at the melting point, respectively). Given that $\Delta \rho < \rho_d$ and that $\alpha_{\rm{eff}} \ll \nu_f$, $v_{\rm{pc}}$ can reasonably be neglected in comparison to other velocities such as, for instance, $u_z$ evaluated at $z=\sqrt{\nu_f t}$, which explains the choice to take $f(0)=f^\prime (0)=0$.

From this analysis, it then becomes possible to evaluate the vertical height $\delta_{\rm{eff}}$ reached by the viscous boundary layer compared with the initial substrate position from $\xi = [\delta_{\rm{eff}}-h(t)]/\sqrt{\nu_f t}$. This yields

\begin{equation}
    \label{delta_eff}
    \delta_{\rm{eff}} = \xi \sqrt{\nu_f t} + s \sqrt{\alpha_{\rm{eff}}t}.
\end{equation}

\noindent Noticeably, this height displays an overall diffusive-like behaviour. Therefore, we introduce an effective water kinematic viscosity, $\nu_{\rm{eff}}$, which is defined as $\delta_{\rm{eff}} \equiv \xi \sqrt{\nu_{\rm{eff}}t}$, so that

\begin{equation}
    \label{nu_eff_ice_1}
    \nu_{\rm{eff}}= \left( \sqrt{\nu_f } + \frac{s}{\xi} \sqrt{\alpha_{\rm{eff}}} \right)^2.
\end{equation}

\noindent In the case of solidification, one obtains $\nu_{\rm{eff}} > \nu_f$, which means that the viscous boundary layer will reach the liquid free surface sooner than for an isothermal drop impact. Freezing thus enhances dissipation, and reduces the spreading diameter. Conversely, when substrate melting occurs $\nu_{\rm{eff}} < \nu_f$: the boundary layer will meet the free surface later than for the isothermal case, so that dissipation appears to be reduced while spreading is favoured. For some experiments, the effective kinematic viscosity that is predicted differs significantly from $\nu_f$: for instance, for drop impacts on ice where $T_s=-2$ \textcelsius\ and $T_d=80$ \textcelsius, one obtains $\nu_{\rm{eff}} \simeq 0.65 \nu_f$. We stress that, in this model, the two cases of liquid solidification and substrate melting during the impact of a droplet onto its solid phase are encompassed into the same framework, which generalizes the approach followed by \citet{2020b_thievenaz}.

\begin{figure}
    \centerline{
    \includegraphics[width=\textwidth]{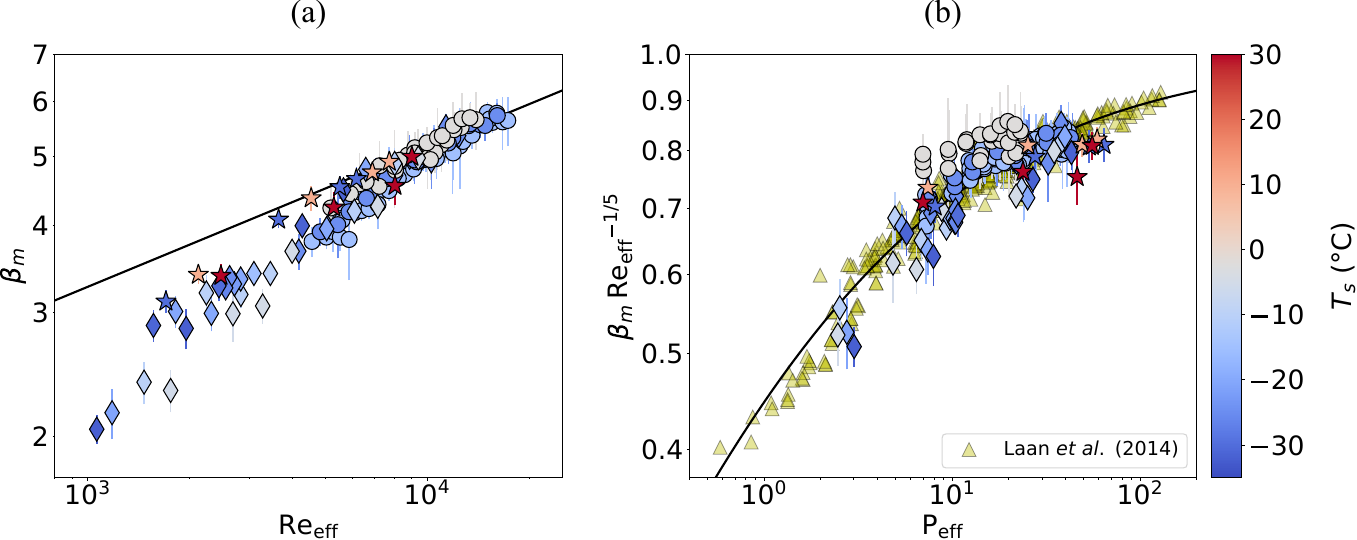}
    }
    \caption{(a) Evolution of $\beta_m$ with the effective Reynolds number $\rm{Re_{eff}} \equiv U D_0/\nu_{\rm{eff}}$, with $\nu_{\rm{eff}}$ the effective kinematic viscosity defined in equation \eqref{nu_eff_ice_1}. The solid line indicates $\beta_m=0.82\, \rm{Re_{eff}} ^{1/5}$. (b) $\beta_m \rm{Re_{eff}}^{-1/5}$ as a function of the impact parameter $\rm{P_{eff}} \equiv \rm{We} \, \rm{Re_{eff}}^{-2/5}$. The data ($\bigtriangleup$) as well as the universal law (solid black line) obtained by \citet{2014_laan} for isothermal drop impacts are also reported. The colourbar indicates the substrate temperature, whereas the symbols correspond to experiments involving ($\lozenge$) water droplets freezing, ($\medcirc$) water droplets causing ice melting, and (\scalebox{0.85}{$\hollowstar$}) DMSO droplets.
    }
    \label{betam_vs_Reeff_Pade_eff}
\end{figure}

As a result of this analysis, we define an effective Reynolds number as $\rm{Re_{eff}} \equiv U D_0/\nu_{\rm{eff}}$, \textit{i.e.}, based on the effective viscosity $\nu_{\rm{eff}}$ that is evaluated by means of equation \eqref{nu_eff_ice_1}. Thus, $\rm{Re_{eff}}$ takes into account the influence of phase change on the development of the viscous boundary layer. The spreading ratio $\beta_m$ is shown as a function of $\rm{Re_{eff}}$ in Figure \ref{betam_vs_Reeff_Pade_eff}(a). This representation reveals a collapse of our experimental data for water onto a master curve, regardless of the nature of the initial substrate (brass or ice) and the dynamics of the ice-water interface (melting or solidification). This shows that the effective Reynolds number better captures the physics at play than the Reynolds number, as highlighted by a comparison with Figure \ref{betam_vs_Re}(a). 


\WS{Nevertheless, a significant number of the present experiments are not compatible with an inertial-viscous scaling of the form $\beta_m \propto \rm{Re_{eff}}^{1/5}$, as highlighted by the comparison between the measured values of $\beta_m$ and the solid line reported in Figure \ref{betam_vs_Reeff_Pade_eff}(a). In addition, data at low $\rm{Re_{eff}}$ appear to be slightly more scattered, and experiments corresponding to DMSO drop impacts (\scalebox{0.85}{$\hollowstar$}) remain above those featuring water droplets freezing ($\lozenge$).} Such a behaviour is reminiscent of the transition from the inertial-viscous regime to the inertial-capillary regime, which has been thoroughly discussed in previous studies \citep{2014_laan,2016_lee}. In the capillary limit, the initial kinetic energy of the drop, which scales as $\rho U^2 {D_0}^3$, is completely converted into surface energy that scales as $\gamma {D_{\rm{min}}}^2$ (with $\rho$ and $\gamma$ the density and surface tension of the drop evaluated at $T_d$, respectively). As a result, one obtains the scaling $\beta_m \sim \rm{We}^{1/2}$, with $\rm{We} = \rho U^2 D_0 / \gamma$ the Weber number \citep{2010_eggers}. Contrariwise, in the viscous regime, the kinetic energy is balanced by viscous dissipation, which leads this time to the scaling $\beta_m \sim \rm{Re}^{1/5}$ \citep{2010_eggers}. To bridge between these two asymptotic scenarios, a universal rescaling has been proposed by \citet{2014_laan} in the context of isothermal drop impacts, in which $\beta_m \rm{Re}^{-1/5}$ is a function of a sole impact parameter, $\rm{P} \equiv \rm{We} \, \rm{Re}^{-2/5}$. Adopting this approach, and using $\rm{Re_{eff}}$ instead of $\rm{Re}$, we plot in Figure \ref{betam_vs_Reeff_Pade_eff}(b) the evolution of the rescaled spreading ratio $\beta_m \rm{Re_{eff}}^{-1/5}$ as a function of the impact parameter $\rm{P_{eff}} \equiv \rm{We} \, \rm{Re_{eff}}^{-2/5}$, for all our experiments. In addition, the data from \citet{2014_laan} ($\bigtriangleup$) corresponding to isothermal drop impacts, are reported in Figure \ref{betam_vs_Reeff_Pade_eff}(b) with $\rm{Re_{eff}}=\rm{Re}$ and $\rm{P_{eff}}=\rm{P}$. Very noticeably, drop impacts involving water solidification ($\lozenge$), DMSO (\scalebox{0.85}{$\hollowstar$}), as well as most experiments with water featuring substrate melting ($\medcirc$) superimpose with the data of \citet{2014_laan}, and are captured by the universal empirical law

\begin{equation}
    \label{pade_law}
    \beta_m \rm{Re_{eff}}^{-1/5} = \frac{\sqrt{\rm{P_{eff}}}}{1.24 + \sqrt{\rm{P_{eff}}}},
\end{equation}

\noindent evidenced by these authors for the isothermal case (solid black line). The typical deviation of these experiments from equation \eqref{pade_law} is less than 10\%, similar to the dispersion of the original data from \citet{2014_laan}. However, a closer inspection reveals that water drop impacts on an ice substrate at $T_s=-2$ \textcelsius\ (grey circles) slightly deviate from relation \eqref{pade_law}. As these experiments belong to the transition region ($\rm{P_{eff}} \sim 10$), and as their initial substrate temperature is close to the melting point, this suggests that wettability effects could start to play a role here \citep{2016_lee}. Nevertheless, as the wetting of water on ice is still a subject of active research, it is not straightforward to conclude on that aspect within the present analysis.

\section{Conclusion}
\label{conclusion}

In the present investigation, experiments of water drop impacts onto ice and cold brass were performed, in which both liquid and substrate temperatures were varied, alongside with the falling height, in order to reach a deeper understanding of the influence of melting and solidification on the impact outcome. The maximum spreading ratio is found to increase with both temperatures as well as with the impact velocity, and the typical size of the corrugations, when present, is proportional to the final radial extent of the main liquid film. Phase change results in a modification of the viscous boundary layer, thereby affecting the overall viscous dissipation occurring within the spreading droplet. Modelling this effect through the use of an effective viscosity allows us to capture the physics at play, and to relate it to a universal law developed for the isothermal configuration. \WS{Additional experiments of dimethyl sulfoxide drop impacts onto a cold brass substrate also show promising agreement with the proposed modelling, which suggests that the approach can be generalized to other fluids.} These results pave the way for a comparison with experiments of molten metal drop impacts on a cold substrate or in the presence of evaporation, which could further validate or enrich the approach followed here. \WS{A detailed study of the effects of frost on drop impacts, using a controlled humidity set-up, would also be needed to reach a better understanding of the environmental situation}. Another configuration of interest, for practical applications as well as to extend the results from the present work, would be to investigate the maximum spreading diameter following non-isothermal drop impacts in the absence of phase change. Indeed, in this scenario, varying the temperature of the initial droplet or the substrate will change the value of the contact temperature, and is thus expected to affect the behaviour of both the thermal and viscous boundary layers. This should, in turn, modify the maximum spreading diameter of the liquid film.


\backsection[Acknowledgements]{The authors warmly thank Caroline Frot and Antoine Garcia for their help in the elaboration of the experimental set-up.}

\backsection[Funding]{This work was partially supported by Agence de l'Innovation de D\'efense (AID) - via Centre Interdisciplinaire d'Etudes pour la D\'efense et la S\'ecurit\'e (CIEDS) - (project 2021 - ICING).}

\backsection[Declaration of interests]{The authors report no conflict of interest.}


\backsection[Data availability statement]{The data that support the findings of this study are available from the corresponding author, upon reasonable request.}

\backsection[Author ORCIDs]{W. Sarlin, https://orcid.org/0000-0002-2668-2279; R. Grivet, https://orcid.org/0000-0002-1489-7336; A. Huerre, https://orcid.org/0000-0003-4702-5128; T. Séon, https://orcid.org/0000-0001-6728-6072; C. Josserand, https://orcid.org/0000-0003-1429-4209}

\backsection[Author contributions]{W. S. and R.G. contributed equally to the present study.}

\appendix


\section{Thermophysical properties of the solid and liquid phases}
\label{appendix_A}

\begin{table}
  \begin{center}
\def~{\hphantom{0}}
  \begin{tabular}{lcccc}
      Substrate & $\rho_s\ \rm{(kg.m^{-3})}$ & $c_{p,s}\ \rm{(J.K^{-1}.kg^{-1})}$ & $k_s\ \rm{(W.m^{-1}.K^{-1})}$ \\[3pt]
      Brass & 8560 & 377 & 121 \\
  \end{tabular}
  \bigskip

  \begin{tabular}{lcccc}
      Solid state & $\rho_i\ \rm{(kg.m^{-3})}$ & $c_{p,i}\ \rm{(J.K^{-1}.kg^{-1})}$ & $k_i\ \rm{(W.m^{-1}.K^{-1})}$ \\[3pt]
      Water (ice) & 916 & 2050 & 2.22 \\
      DMSO & 1104.7 & 1912 & 0.174 \\
  \end{tabular}
  \bigskip

  \begin{tabular}{lcccccc}
      Liquid state & $T_f\ \rm{(K)}$ & $\mathcal{L}_f\ \rm{(kJ.kg^{-1})}$ & $\rho_d\ \rm{(kg.m^{-3})}$ & $c_{p,d}\ \rm{(J.K^{-1}.kg^{-1})}$ & $k_d\ \rm{(W.m^{-1}.K^{-1})}$ \\[3pt]
      Water & 273.15 & 333 & 999.8 & 4219.9 & 0.556 \\
      DMSO & 291.65 & 172.9 & 1095.5 & ~~1960~ & 0.174 \\
  \end{tabular}
  \caption{Thermal properties of the solid and liquid phases involved in the present study. $T_f$ corresponds to the melting point, $\mathcal{L}_f$ to the latent heat of fusion, and $\rho$, $c_p$, and $k$ are the density, specific heat, and thermal diffusivity, respectively (with their associated subscripts $s$, $i$, or $d$ denoting the substrate, the ice, or the liquid, respectively.}
  \label{tab:tab1}
  \end{center}
\end{table}

Table \ref{tab:tab1} reports the thermal properties of the different solids and liquids involved in the present study. The values of densities, specific heats, and thermal conductivities indicated are those used when solving the Stefan problem (see Appendix \ref{appendix_C}) to obtain equation \eqref{transcendental_equation} in the main text. From these quantities, one can readily obtain the thermal diffusivity $\alpha_n=k_n/(\rho_n c_{p,n})$ and effusivity $r_n=\sqrt{k_n \rho_n c_{p,n}}$, where subscript $n$ corresponds either to $s$, $i$, or $d$ depending on whether the substrate, the ice, or the liquid is to be considered, respectively.

As stated in the main text, some quantities are also evaluated at the melting point or at the droplet temperature, $T_d$. To this end, and for water, standard correlations have been used to estimate densities \citep{2009_patek}, viscosities \citep{2009_patek,2015_dehaoui}, and surface tensions \citep{2016_patek}. For dimethyl sulfoxide, for which $T_d=25$\textcelsius, droplet density was taken as $\rho=1095.5\ \rm{kg.m^{-3}}$, kinematic viscosity as $\nu_f=1.83\times10^{-6}\ \rm{m^2.s^{-1}}$, and surface tension as $\gamma=43.53\ \rm{mN.m^{-1}}$.

%
%

\section{Temperature loss during the droplet's fall}
 \label{appendix_B}

For the purpose of this investigation, it is important to estimate the temperature loss of the droplet over the course of its fall. To do so, assuming the droplet temperature $T_w$ to be uniform spatially, we observe that its variation with time reads

\begin{equation}
\label{temperature_drop_1}
m c_{p,d} \frac{dT_w}{dt}=\pi C(t){D_0}^2 \left( T_w(t) - T_a \right) \leqslant \pi C(t){D_0}^2 \left( T_{d} - T_a \right),
\end{equation}

\noindent where $D_0$ is the initial diameter of the drop, whose shape is assumed not to change with time, $m$ is its mass, $T_d$ and $T_a$ are the initial temperatures of the liquid droplet and the air, respectively, $c_{p,d}$ is the specific heat of the liquid considered, and $C$ is the convective heat exchange coefficient. The right-hand side of equation \eqref{temperature_drop_1} constitutes an upper bound of the actual temperature loss. Then, one may note that $C(t)$ can be expressed as $\mathrm{Nu}\, k_a/D_0$, with $k_a$ the thermal conductivity of the air and $\mathrm{Nu}$ the Nusselt number. Integrating \eqref{temperature_drop_1} over time, and making the conservative assumption of a pure free fall thereby leads to the following inequality for the global temperature loss $\Delta T$

\begin{equation}
\label{temperature_drop_2}
\Delta T \leqslant \frac{6 \left( T_d - T_a \right) k_a}{\rho c_{p,d} {D_0}^2} \int_0^{t_f} \mathrm{Nu}(t) \, \odif{t},
\end{equation}

\noindent where $t_f=\sqrt{2H/g}$ corresponds to the free-fall time over a vertical distance $H$ (with $g$ the gravitational acceleration), and $\rho$ to the liquid density at $T_w=T_d$. From there, it is possible to use classical correlations to express the Nusselt number, such as those established by \citet{1960_yuge} for a sphere in forced convection, yielding

\begin{gather}
\label{yuge_correlations}
\mathrm{Nu} = 2+0.493\,{\mathrm{Re}_a}^{0.5}\ \left( 10 < \mathrm{Re}_a < 1.8 \times 10^3 \right),\\
\notag \\
\mathrm{Nu} = 2+0.3\,{\mathrm{Re}_a}^{0.5664}\ \left( 1.8 \times 10^3 < \mathrm{Re}_a < 1.5 \times 10^5 \right),
\end{gather}

\noindent where $\mathrm{Re}_a=U(t)D_0/\nu_a$ is the Reynolds number associated to the air motion, with $U(t)$ and $\nu_a$ the droplet velocity at time $t$ and the air kinematic viscosity, respectively.

\noindent Finally, by defining $\mathrm{Re}_c=1.8 \times 10^3$ and $t_c=\nu_a \mathrm{Re}_c/(gD_0)$, one gets the following upper bound for the temperature loss $\Delta T$ during the droplet's fall

\begin{eqnarray}
\label{temperature_drop_3}
\notag
\Delta T \leqslant \frac{6 \left( T_d - T_a \right) k_a}{\rho_w c_{p,d} {D_0}^2} & \Big\{ & 2\, t_f + \frac{0.986}{3} \left( \frac{gD_0}{\nu_a} \right)^{0.5} {t_c}^{3/2} \\
 & + & \frac{0.3}{1.5664} \left( \frac{gD_0}{\nu_a} \right)^{0.5664} \left( {t_f}^{1.5664} - {t_c}^{1.5664} \right) \Big\}.
\end{eqnarray}

If, then, the left-hand side of equation \eqref{temperature_drop_3} is evaluated for water and for the various initial parameters explored in the present investigation, it is found that the maximal temperature loss (occurring for $T_d=80$ \textcelsius\ and $H=2.2$ m) is less than $2$ \textcelsius. In the case of DMSO drop impacts, for which $T_d$ is very close to $T_a$, the estimated change is below 1 \textcelsius. To evaluate equation \eqref{temperature_drop_3}, the following values for $T_a$ have been taken: $T_a=25$ \textcelsius\ in the case of water and $T_a=18$ \textcelsius\ for DMSO. Consequently, we choose to neglect this temperature change over the course of the present study for the sake of simplicity.

\section{The three phase Stefan Problem}
\label{appendix_C}

\begin{figure}
    \centerline{
    \includegraphics[width=0.55\textwidth]{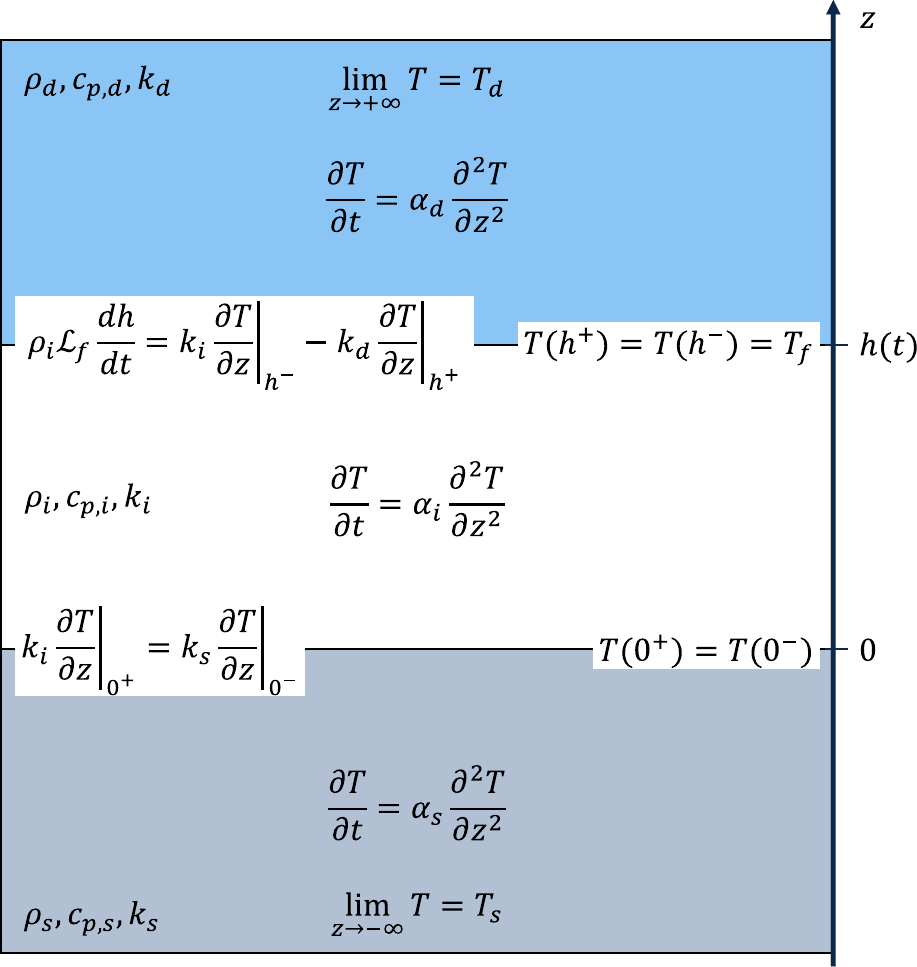}
    }
    \caption{\WS{Model equations and boundary conditions for the three phase Stefan problem involving the liquid phase (expanding droplet), the growing ice layer, and the solid substrate.}}
    \label{three_phase_stefan_problem}
\end{figure}

The Stefan problem consists of solving the heat diffusion equation in several phases, in the presence of a moving interface of phase change \citep{1971_rubinstein}. In the present study, we are interested in the one-dimensional formulation of this problem involving either three phases for solidification (the liquid state, its corresponding ice, and the solid substrate), or two for melting (the liquid state and its ice). The first situation is detailed below, and is strongly inspired by the self-similar analysis performed by \citet{2019b_thievenaz}. The model equations and boundary conditions are summarized in figure \ref{three_phase_stefan_problem} for the three phase Stefan problem involving the liquid phase (droplet), the growing ice layer, and the solid substrate. The properties for each phase are respectively denoted by the subscript letters $d$, $i$, and $s$. Assuming that advection can be neglected in the problem (see main text), the equations to be solved in the different phases are

\begin{align}
\label{b_11}
\partial_t T & = \alpha_d \partial_z^2 T & z > h(t), \\
\label{b_12}
\partial_t T & = \alpha_i \partial_z^2 T & 0 < z < h(t), \\
\label{b_13}
\partial_t T & = \alpha_s \partial_z^2 T & z < 0,
\end{align}

\noindent where $T$ is the temperature field, and $\alpha_d$, $\alpha_i$, and $\alpha_s$ the thermal diffusivities of the different phases ($\alpha_n=k_n/(\rho_n c_{p,n})$, where subscript $n$ corresponds either to $s$, $i$, or $d$ depending on whether the substrate, the ice, or the liquid is to be considered, respectively). In addition, the relevant boundary conditions to apply read

{\allowdisplaybreaks
\begin{eqnarray}
\label{b_21}
\lim_{z \to +\infty} T & = & T_d, \displaystyle \\
\label{b_22}
T(h^-) = T(h^+) & = & T_f, \\
\label{b_23}
\rho_i \mathcal{L}_f \rm{d}_t h & = & k_i \left. \partial_z T \right|_{h^-} - k_d \left. \partial_z T \right|_{h^+}, \\
\label{b_24}
T(0^-) & = & T(0^+), \\
\label{b_25}
k_s \left. \partial_z T \right|_{0^-} & = & k_i \left. \partial_z T \right|_{0^+}, \\
\label{b_26}
\lim_{z \to -\infty} T & = & T_s, \displaystyle
\end{eqnarray}
}

\noindent with $T_d$ and $T_s$ the droplet and substrate temperatures, respectively, $T_f$ the melting point, and $h$ the position of the moving liquid-ice interface. In addition, $k_s$, $k_i$, and $k_d$ stand for the thermal conductivities of the substrate, the ice and of the liquid phase, respectively, $\rho_i$ is the ice density, and $\mathcal{L}_f$ the latent heat of fusion. In particular, condition \eqref{b_22} ensures that the temperature at the liquid-ice interface corresponds to the melting point, equation \eqref{b_25} establishes the equality of thermal fluxes at the ice-substrate interface ($z=0$), and expression \eqref{b_23} is the so-called Stefan condition. It expresses that the growth of the liquid-ice interface is dictated by the local difference of thermal fluxes (at $z=h$).

One may then apply the transformations $z \equiv H \, \overline{z}$, $h \equiv H \, \overline{h}$, $t \equiv (H^2/\alpha_i) \, \overline{t}$ and $T \equiv T_s + (T_f-T_s) \, \overline{T}$ to equations \eqref{b_11}-\eqref{b_13} and \eqref{b_21}-\eqref{b_26} to obtain the dimensionless problem

\begin{align}
\label{b_31}
\partial_{\overline{t}} \overline{T} & = \omega_d \partial_{\overline{z}}^2 \overline{T} & \overline{z} > \overline{h}(\overline{t}), \\
\label{b_32}
\partial_{\overline{t}} \overline{T} & = \partial_{\overline{z}}^2 \overline{T} & 0 < \overline{z} < \overline{h}(\overline{t}), \\
\label{b_33}
\partial_{\overline{t}} \overline{T} & = \omega_s \partial_{\overline{z}}^2 \overline{T} & \overline{z} < 0,
\end{align}

\noindent with $\omega_d=\alpha_d/\alpha_i$, $\omega_s=\alpha_s/\alpha_i$, and with the associated boundary conditions

{\allowdisplaybreaks
\begin{eqnarray}
\label{b_41}
\lim_{\overline{z} \to +\infty} \overline{T} = \overline{T}_d & = & \frac{T_d-T_s}{T_f-T_s}, \displaystyle \\
\label{b_42}
\overline{T}(\overline{h}^-) = \overline{T}(\overline{h}^+) & = & 1, \\
\label{b_43}
\frac{1}{\mathrm{St}} \rm{d}_{\overline{t}} \overline{h} & = & \left. \partial_{\overline{z}} \overline{T} \right|_{\overline{h}^-} - \kappa \left. \partial_{\overline{z}} \overline{T} \right|_{\overline{h}^+}, \\
\label{b_44}
\overline{T}(0^-) & = & \overline{T}(0^+), \\
\label{b_45}
k_s \left. \partial_{\overline{z}} \overline{T} \right|_{0^-} & = & k_i \left. \partial_{\overline{z}} \overline{T} \right|_{0^+}, \\
\label{b_46}
\lim_{\overline{z} \to -\infty} \overline{T} & = & 0, \displaystyle
\end{eqnarray}
}

\noindent with $\mathrm{St}=c_{p,i} (T_f-T_s) / \mathcal{L}_f$ the Stefan number, whose definition involves the heat capacity $c_{p,i}=k_i/(\rho_i \alpha_i)$ of the ice, and $\kappa = k_d / k_i$. The Stefan number is a dimensionless quantity that compares the energy needed to decrease the ice temperature from $T_f$ to $T_s$ to the energy released by phase change, both energies being considered per unit mass.

It is possible to obtain a self-similar solution for equations \eqref{b_31}-\eqref{b_33} that satisfies boundary conditions \eqref{b_41}-\eqref{b_46} by conducting a similar analysis to that presented in \citet{2019a_thievenaz}. By doing so, a self-similar temperature field can be found only if it solely depends on the variable $\eta = \overline{z}/\sqrt{\overline{t}}$, whereas the liquid-ice front position $\overline{h}$ is proportional to $\sqrt{\overline{t}}$. This solution reads

\begin{align}
\label{b_51}
\overline{T} (\eta) & = \overline{T}_d + \frac{\overline{T}_d-1}{1-\erf \left( \chi/(2 \sqrt{\omega_d}) \right)} \left( \erf \left( \eta/(2 \sqrt{\omega_d}) \right) - 1 \right) & \overline{z} > \overline{h}(\overline{t}), \\
\label{b_52}
\overline{T} (\eta) & = \frac{1}{r_i/r_s + \erf \left( \chi / 2 \right)} \left( \frac{r_i}{r_s} + \erf \left( \eta / 2 \right) \right) & 0 < \overline{z} < \overline{h}(\overline{t}), \\
\label{b_53}
\overline{T} (\eta) & = \frac{r_i/r_s}{r_i/r_s + \erf \left( \chi / 2 \right)} \left( 1 + \erf \left( \eta / \left( 2 \sqrt{\omega_s} \right) \right) \right) & \overline{z} < 0,
\end{align}

\noindent where $r_i=k_i/\sqrt{\alpha_i}$ and $r_s=k_s/\sqrt{\alpha_s}$ are the thermal effusivities of the ice and the substrate, respectively, while $\chi$ is related to the position of the liquid-solid moving interface through $\overline{h}(\overline{t}) = \chi \sqrt{\overline{t}}$ and is solution to the following transcendental equation 

\begin{equation}
    \label{b6}
    \frac{\chi \sqrt{\pi}}{2 \rm{St}} = \frac{e^{-\chi^2/4}}{r_i/r_s+\erf \left( \chi/2 \right)} + \frac{r_d}{r_i} \frac{e^{-\chi^2/(4 \omega_d)}}{1-\erf \left( \chi/(2\sqrt{\omega_d}) \right)} \frac{T_f-T_d}{T_f-T_s},
\end{equation}

\noindent with $r_d=k_d/\sqrt{\alpha_d}$ the thermal effusivity of water.


\bibliographystyle{jfm}
\bibliography{bibliography}

\end{document}